\documentclass[a4paper, DIV=12, BCOR=0pt]{scrartcl}

\usepackage[utf8]{inputenc}   
\usepackage[english]{babel}
\usepackage{csquotes}
\usepackage[hidelinks]{hyperref}

\usepackage{xcolor}

\colorlet{deviceBoundary}{gray!70!black}
\colorlet{deviceFill}{gray}

\usepackage{graphicx}
\usepackage{epsfig}
\usepackage{booktabs}
\usepackage{wrapfig2}

\usepackage[exponent-product = \cdot]{siunitx}
\usepackage{amsfonts,amsmath,amssymb,array} 
\input{cmrfont.sty}           

\usepackage{tikz}
\usetikzlibrary{matrix, positioning, fit, calc, shapes}

\newcommand{\Abstract}{
	Molecular communication is a novel approach for data transmission between miniaturised devices, especially in contexts where electrical signals are to be avoided. The communication is based on sending molecules (or other particles) at nanoscale through a typically fluid channel instead of the ``classical'' approach of sending electrons over a wire. 
	
	Molecular communication devices have a large potential in future medical applications as they offer an alternative to antenna-based transmission systems that may not be applicable due to size, temperature, or radiation constraints.
		The communication is achieved by transforming a digital signal into concentrations of molecules that represent the signal. These molecules are then detected at the other end of the communication channel and transformed back into a digital signal. Accurately modeling the transmission channel is often not possible which may be due to a lack of data or time-varying parameters of the channel (e.\,g., the movements of a person wearing a medical device). This makes the process of demodulating the signal (i.\,e., signal classification) very difficult.
	
	Many approaches for demodulation have been discussed in the literature with one particular approach having tremendous success -- artificial neural networks. These artificial networks imitate the decision process in the human brain and are capable of reliably classifying even rather noisy input data. Training such a network relies on a large set of training data. As molecular communication as a technology is still in its early development phase, this data is not always readily available. In this paper, we discuss neural network-based demodulation approaches relying on synthetic simulation data based on theoretical channel models as well as works that base their network on actual measurements produced by a prototype test bed.
	
	In this work, we give a general overview over the field molecular communication, discuss the challenges in the demodulations process of transmitted signals, and present approaches to these challenges that are based on artificial neural networks.
}

\title{Artificial Intelligence for Molecular Communication}
\date{01.05.2023}
\author{Max Bartunik \href{mailto:max.bartunik@fau.de}{\texttt{max.bartunik@fau.de}}\\
	 Jens Kirchner \href{mailto:jens.kirchner@fau.de}{\texttt{jens.kirchner@fau.de}}\\
	  Oliver Keszocze \href{mailto:oliver.keszoecze@fau.de}{\texttt{oliver.keszoecze@fau.de}}}

\newcommand{\authoronename}{Max Bartunik}
\newcommand{\authoronecv}{completed his master's degree for electrical engineering in 2019 at the Friedrich-Alexander-Universität Erlangen-Nürnberg (FAU). He is currently pursuing his Ph.D. as a research assistant at the Institute for Electronics Engineering of the FAU. His main research topics are molecular communication and medical electronics.}
\newcommand{\authoroneaddress}{Friedrich-Alexander-Universität Erlangen-Nürnberg (FAU), Lehrstuhl für Technische Elektronik
	\mbox{E-Mail}:~m\href{mailto:max.bartunik@fau.de}{\texttt{max.bartunik@fau.de}}}
\newcommand{\authoronephotopath}{bartunik.jpg}

\newcommand{\authortwoname}{Jens Kirchner}
\newcommand{\authortwocv}{studied physics at the Friedrich-Alexander-Universität Erlangen-Nürnberg (FAU) and at University of St. Andrews, Scotland. He received his doctorates in 2008 and 2016 from FAU in the fields of biosignal analysis and philosophy of science to the Dr. rer. nat. and Dr. phil, respectively. Between 2008 and 2015 he worked at Biotronik SE \& Co. KG in Erlangen and Berlin in the research and development of implantable cardiac sensors. Since 2015, he is with the Institute for Electronics Engineering at FAU, where he heads the Medical Electronics \& Multiphysics Systems group. His research interests lie in wearable and implantable sensors, inductive power transfer, and molecular communication. He is a Senior Member of the IEEE with membership in the Communications Society, the Magnetics Society and the Engineering in Medicine and Biology Society.}
\newcommand{\authortwoaddress}{Friedrich-Alexander-Universität Erlangen-Nürnberg (FAU), Lehrstuhl für Technische Elektronik \mbox{E-Mail}:~\href{mailto:jens.kirchner@fau.de}{\texttt{jens.kirchner@fau.de}}}
\newcommand{\authortwophotopath}{kirchner.jpg}

\newcommand{\authorthreename}{Oliver Keszöcze}
\newcommand{\authorthreecv}{is with the Friedrich-Alexander-Universität Erlangen-Nürnberg (FAU), Germany where he is a Junior Professor at the Chair for Hardware/Software-Co-Design since 2018.
	He received a Diploma degree in applied mathematics and a B.Sc. in Computer Science from the University of Bremen in 2011.
	After a short career as a Software Engineer, he decided to pursue a Ph.D. in Computer Science in 2012 at his Alma Mater.
	Since 2014 he was also a Researcher with the German Research Center for Artificial Intelligence.
	In 2017,  he received his doctoral degree (Dr. rer. nat) at the University of Bremen.
	
	Prof. Keszocze's research interests include several aspects of Logic Synthesis and Computer Aided Design in various application domains and for different technologies.
	His current research focuses on Approximate Computing for both, ASICs and FPGAs.
	
	He has been serving as a TPC member for several conferences, including DATE, ICCAD, and ASP-DAC and is a reviewer for IEEE TCAD as well as for several other journals.
	He is the organizer of the annual Special Session ``Future Trends in Emerging Technologies'' at the DSD Conference.
	He is glad to be able to support young researchers by being part of DATE conference's PhD Forum TPC.
}
\newcommand{\authorthreeaddress}{Friedrich-Alexander-Universität Erlangen-Nürnberg (FAU), Hardware-Software-Co-Design \mbox{E-Mail}:~\href{mailto:oliver.keszoecze@fau.de}{\texttt{oliver.keszoecze@fau.de}}}
\newcommand{\authorthreephotopath}{keszocze_cropped_small.jpg}

\newcommand{\authorone}{\noindent\textbf{\authoronename} \authoronecv
	
	\medskip
	\noindent Address: \authoroneaddress}
\newcommand{\authortwo}{\noindent\textbf{\authortwoname} \authortwocv
	
	\medskip\noindent
	Address: \authortwoaddress}
\newcommand{\authorthree}{\noindent\textbf{\authorthreename} \authorthreecv
	
	\medskip\noindent
	Address: \authorthreeaddress}

\usepackage{todonotes}
\usepackage{siunitx}
\usepackage{pgf}
\usepackage{pgfplots}
\usepackage{pgfplotstable}
\usepackage{circuitikz}
\tikzset{meter/.append style={draw, inner sep=10, rectangle, font=\vphantom{A}, minimum width=30, line width=.8,
		path picture={\draw[black] ([shift={(.1,.3)}]path picture bounding box.south west) to[bend left=50] ([shift={(-.1,.3)}]path picture bounding box.south east);\draw[black,-latex] ([shift={(0,.1)}]path picture bounding box.south) -- ([shift={(.3,-.1)}]path picture bounding box.north);}}}
\usetikzlibrary{pgfplots.colormaps, calc, shapes, fit}
\usepgfplotslibrary{colorbrewer}
\pgfplotsset{
	cycle list/Set1,
	cycle multiindex* list={
		mark list*\nextlist
		Set1\nextlist
	},
	colormap={confusion}{
		rgb255=(255,255,255)
		rgb255=(208,209,230)
		rgb255=(166,189,219)
		rgb255=(116,169,207)
		rgb255=(54,144,192)
		rgb255=(5,112,176)
		rgb255=(3,78,123)
	}
}
\newcommand{\whitethreshold}{0.65} 

\DeclareMathOperator*{\argmin}{arg\,min}

\usetikzlibrary{pgfplots.colormaps, calc, shapes, fit}
\usetikzlibrary{decorations.pathmorphing}

\newcommand*{\ReadOutElement}[4]{%
	\pgfplotstablegetelem{#2}{[index]#3}\of{#1}%
	\let#4\pgfplotsretval
}

\usepackage{pifont}
\newcommand{\cmark}{\ding{51}}%
\newcommand{\xmark}{\ding{55}}%

\usepackage{bm}

\usepackage{overpic}

\newcommand{\diff}[1]{#1}

\begin{document}

\maketitle

\begin{abstract}
\Abstract

	\par\vskip\baselineskip\noindent
\end{abstract}

\section{Introduction}
Molecular communication (MolCom) is a concept that has been discussed in the literature for quite some time (\diff{\cite{Nakano:2005}, see also} \cite{NakanoEckford:2013, Nakano2010, Pierobon2013}). The first physical implementation of the concept was presented in~\cite{Farsad2013}. Since then, multiple technical implementations using very different means of sending data, ranging from biological setups using bacteria \cite{Grebenstein2019} to fluorescent jets in air \cite{Damrath2021}, have been proposed (e.g., air-based or fluid-based setups). These implementations vary in their application domain and technical details such as transmission speed and quality \diff{(see, e.g., \cite{Lotter:2023:1, Lotter:2023:2} for an overview)}.

\begin{figure*}[b]
	\centering
	\begin{tikzpicture}[
		device/.style={rectangle,draw=deviceBoundary, thick, fill=deviceFill, align=center, text=white, text centered},
		octagon/.style={regular polygon, regular polygon sides=8}
		]
		\node[device,text width=2.5cm, minimum height=1cm] (trans) {Transmitter};
		\node[left=15pt of trans, pin={[pin edge={<-}]100:$S$}] (s) {\verb|"message"|};
		\node[device,text width=2.5cm, minimum height=1cm, right=3cm of trans] (recv) {{Receiver}};
		
		\draw[->] (s) -- (trans);
		
		\node[right=0pt of recv, pin={[pin edge={<-}]100:$S'$}] (s') {\verb|"message"|};
		
		\draw[deviceBoundary,thick] ($(trans.east)+(0,.3)$) -- ($(recv.west)+(0,.3)$);
		\draw[deviceBoundary,thick] ($(trans.east)-(0,.3)$) -- ($(recv.west)-(0,.3)$);
		
		\node[fill=blue!50!white, circle, scale=0.5] at ($(trans.east)!.55!(recv.west)+(0,-.05)$) {};
		\node[fill=blue!50!white, circle, scale=0.6] at ($(trans.east)!.2!(recv.west)+(0,.15)$) {};
		\node[fill=blue!50!white, circle, scale=0.3] at ($(trans.east)!.8!(recv.west)+(0,-.04)$) {};
		\node[fill=blue!50!white, circle, scale=0.5] at ($(trans.east)!.05!(recv.west)+(0,-.1)$) {};
		\node[fill=blue!50!white, circle, scale=0.7] at ($(trans.east)!.95!(recv.west)+(0,-.16)$) {};
		
		\node[fill=blue!70!white, octagon, scale=0.5, rotate=20] at ($(trans.east)!.4!(recv.west)+(0,.0)$) {};
		\node[fill=blue!70!white, octagon, scale=0.2] at ($(trans.east)!.9!(recv.west)+(0,.2)$) {};
		\node[fill=blue!70!white, octagon, scale=0.7] (molecule) at ($(trans.east)!.7!(recv.west)+(0,.1)$) {};
		\node[fill=blue!70!white, octagon, scale=0.5] at ($(trans.east)!.25!(recv.west)+(0,-.15)$) {};

		%
		%
		
		\draw[<-] ($(trans.east)+(0,.3)!.1!(recv.west)+(0,.1)$) -- node[above, pos=1] {channel} + (-.2,.2);
		\draw[<-] (molecule) -- node[above, pos=1] {molecule} +(.5,.5);
		\draw[<-] ($(trans.east)!.47!(recv.west)+(0,.15)$) -- node [above, outer sep=.1pt, pos=1] {transport medium (e.g., air or fluid)} +(0,1.3);
	\end{tikzpicture}
	
	\caption{General molecular communication setup: A transmitter converts an incoming signal $S$ into a stream of molecules (or other particles) in the nanoscale. The molecules are then transported through a channel (filled with, e.g., air or fluid, see Sec.~\ref{sec:air-based} and~\ref{sec:fluid-based}, respectively) to a receiver that tries to recover the initial signal $S$ in its output signal $S'$. See Sec.~\ref{sec:ai} for details on the recovery process.}
	\label{fig:concept}
\end{figure*}
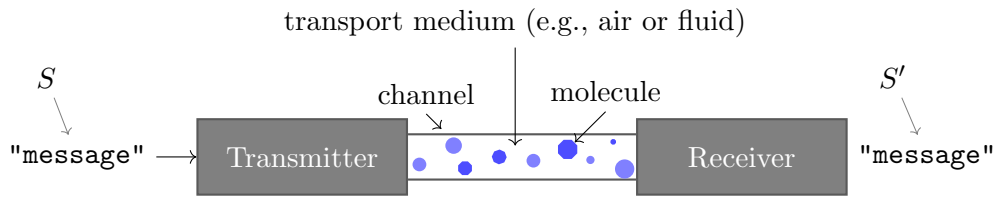

The general setup of a MolCom transmission channel is depicted in Fig.~\ref{fig:concept}. The main idea is to take a given input signal $S$ and modulate it in such a way that it can be represented by molecules (or possibly other small particles), i.\,e., information particles, to be sent through a transmission channel filled with some transport medium like air or a fluid\footnote{While molecular communication is technically possible using solid bodies as the transmission channel, it doesn't have any practical applications so far and, hence, is not covered in this article.}. Within the field of MolCom, the smallest amount of information that can be sent (i.\,e., an individual element of the modulation alphabet) is referred to as \emph{symbol}. Nevertheless, in order to allow for comparability between different MolCom implementations as well as to conventional communication, based on electromagnetic waves, the transmission speed is often given in bits per second (\unit{\bit\per\second}).

In the following, we will shortly present two different MolCom implementations, one based on air as the transport medium and one using fluids before discussing the problem, demodulation, that will be solved using artificial intelligence. 

\subsection{Air-based MolCom Setups}\label{sec:air-based}

\begin{figure}
	\centering
	\includegraphics[width=.65\linewidth]{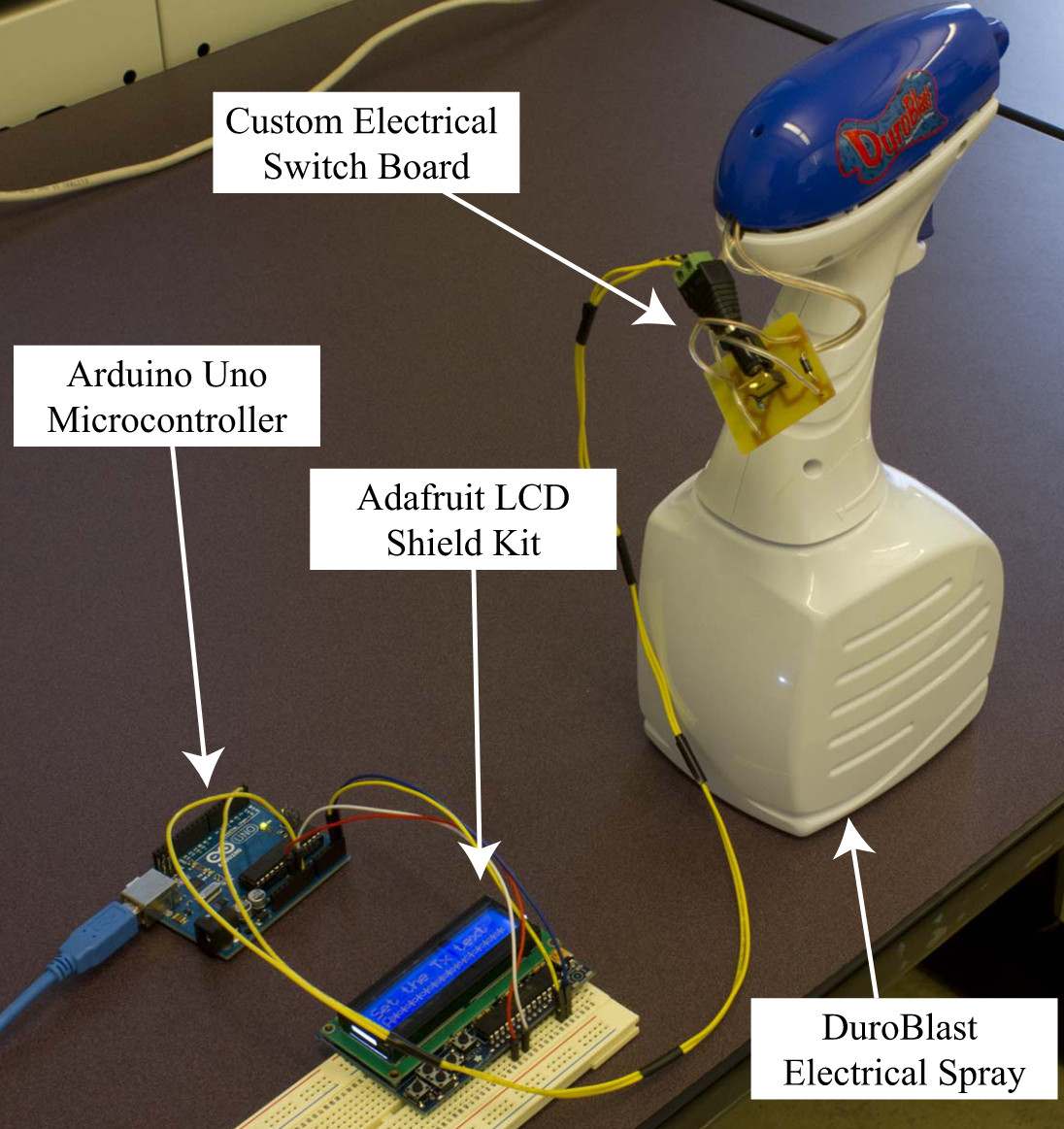}
	\caption{Air-based MolCom transmitter setup: An Arduino microcontroller is used to control alcohol emissions from a prepared plastic spray can. (Image taken from \cite[Fig.~2]{Farsad2013})}
	\label{fig:air_transmitter}
\end{figure}

One of the first proof-of-concepts for MolCom was presented in~\cite{Farsad2013}. The authors used air as the transport medium and alcohol molecules to encode the signal to be sent. Their transmitter is shown in Fig.~\ref{fig:air_transmitter}. The setup only allows for a transmission speed of \SI{0.33}{\bit\per\second}. This proof-of-concept was intended to serve as a starting point for further research. Multiple papers have been published on air-based MolCom since. The setup of~\cite{Lu2017}, for example, reaches a transmission speed of up to \SI{2}{\bit\per\second}. An even higher speed of up to \SI{40}{\bit\per\second} has been reported in~\cite{Damrath2021}. It should be noted, though, that in order to reach that speed, the authors restricted the air movement by spraying particles in a tube of fixed size and well-defined light settings. This makes it rather difficult to apply this kind of communication to specific applications.

In general, not having a fixed channel to directly send the molecules through (after an initial acceleration, the molecules float in the air) reduces the applicability of air-based MolCom. One domain of interest might be the modeling of spreading of infections~\cite{Hoeher2022}.

\subsection{Fluid-based MolCom Setups}\label{sec:fluid-based}

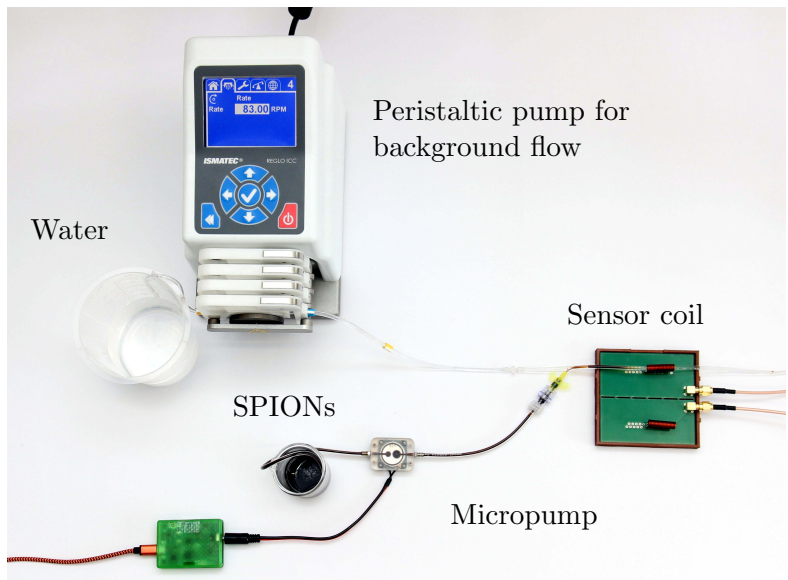
\begin{figure}
	\centering
	\begin{overpic}[width=0.65\linewidth]{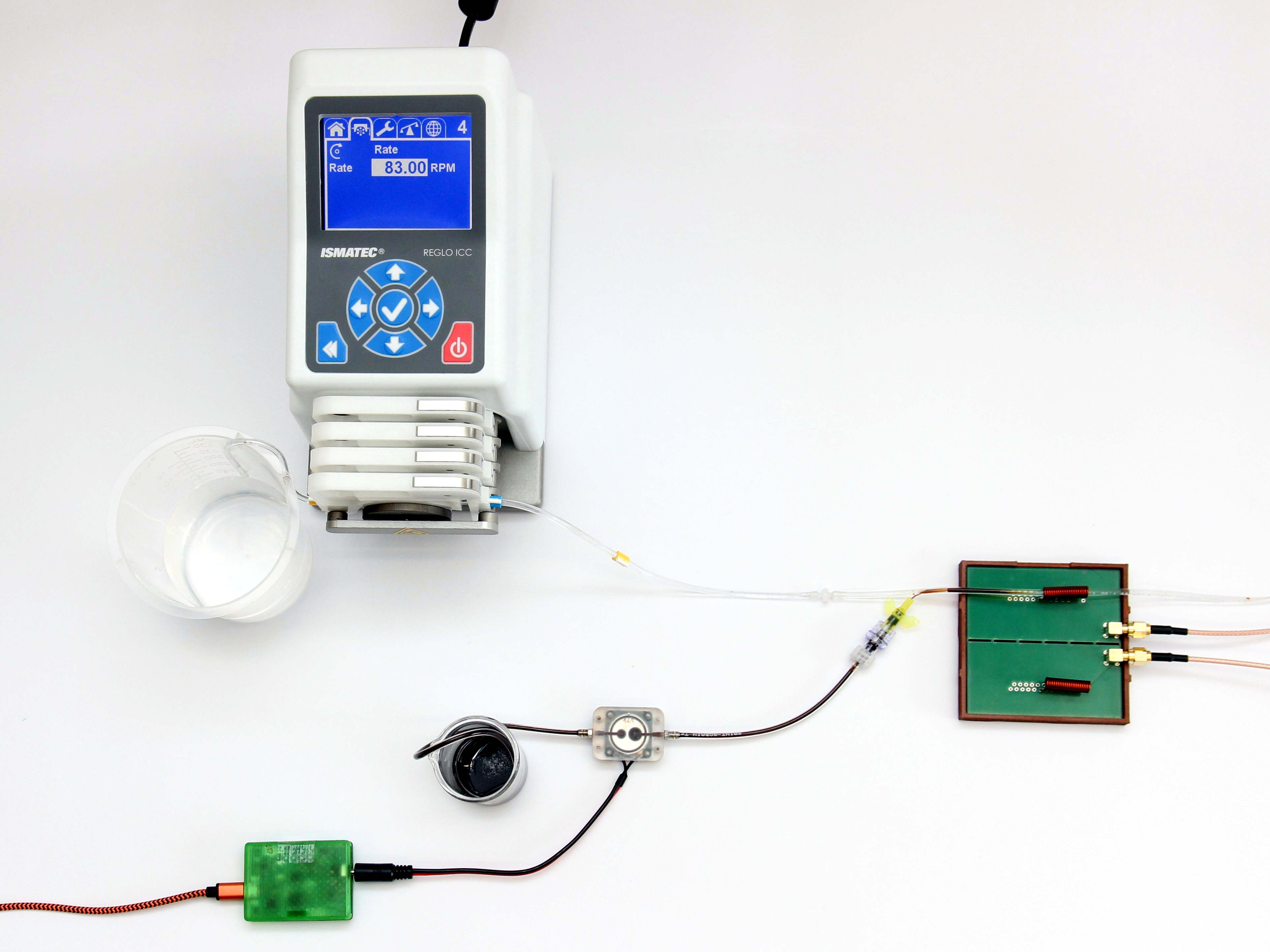}
		\put(47,58){\parbox{100pt}{Peristaltic pump for\\background flow}}
		\put(3,45){Water}
		\put(57,8){Micropump}
		\put(29,22){SPIONs}
		\put(72,34){Sensor coil}
	\end{overpic}
	\caption{Fluid-based molecular communication setup: A peristaltic pump provides the background flow of the transport fluid. A micropump ejects SPIONs that are then detected by a sensor coil. (Image adapted from~\cite[Fig.~8]{Bartunik2023})}
	\label{fig:fluid_molcom_setup}
\end{figure}

In contrast to air-based MolCom, fluid-based setups make use of a dedicated physical channel that hosts the transport medium (e.g., water) and moves the molecules from the transmitter to the receiver along a well-defined path. Figure~\ref{fig:fluid_molcom_setup} shows an exemplary fluid-based MolCom setup. A peristaltic pump provides a constant background flow of the transport medium along a tube. A micropump is used to inject superparamagnetic iron-oxide nanoparticles (SPIONs) into the tube. At the end of the tube, the concentration of SPIONs is measured using a sensor coil.

Fluid-based MolCom can be applied in situations where an existing physical channel can be used and additional electrical wires are not feasible, e.g., due to high installation cost or safety concerns. Specifically promising use-cases for fluid-based MolCom can be found in the medical field, with applications within the (human) body.

\subsection{Challenge for MolCom: Demodulation}
Fig.~\ref{fig:concept} visualizes the transmitter and receiver as black boxes without showing any internals. To perform their operations, both devices actually have to conduct multiple steps, including the step of interest in this paper: demodulation. Hence, we will shortly review how a receiver for MolCom works and then discuss the problems of demodulation.

While the following discussion uses fluid-based MolCom to illustrate the challenge at hand, the general issue also holds for other types of MolCom, e.g., air-based MolCom as discussed in Sec.~\ref{sec:air-based}.

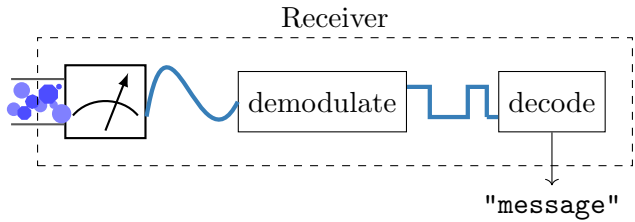
\begin{figure}
	\centering
	\begin{circuitikz}[
		device/.style={rectangle,draw=deviceBoundary, thick, fill=deviceFill, align=center, text=white, text centered},
		octagon/.style={regular polygon, regular polygon sides=8}
		]
		\node[minimum height=1cm] (trans) {};
		\node[meter, minimum height=.8cm, align=center, right=.7cm of trans, draw] (demod) {};
		\node[minimum height=.8cm, align=center, right=1.2cm of demod, draw] (dmod) {demodulate};
		\node[minimum height=.8cm, align=center, right=1.2cm of dmod, draw] (decode) {decode};
		
		\draw[deviceBoundary,thick] ($(trans.east)+(0,.3)$) -- ($(demod.west)+(0,.3)$);
		\draw[deviceBoundary,thick] ($(trans.east)-(0,.3)$) -- ($(demod.west)-(0,.3)$);
		
		\node[fill=blue!50!white, circle, scale=0.5] at ($(trans.east)!.55!(demod.west)+(0,-.05)$) {};
		\node[fill=blue!50!white, circle, scale=0.6] at ($(trans.east)!.2!(demod.west)+(0,.15)$) {};
		\node[fill=blue!50!white, circle, scale=0.3] at ($(trans.east)!.8!(demod.west)+(0,-.04)$) {};
		\node[fill=blue!50!white, circle, scale=0.5] at ($(trans.east)!.05!(demod.west)+(0,-.1)$) {};
		\node[fill=blue!50!white, circle, scale=0.7] at ($(trans.east)!.95!(demod.west)+(0,-.16)$) {};
		
		\node[fill=blue!70!white, octagon, scale=0.5, rotate=20] at ($(trans.east)!.4!(demod.west)+(0,.0)$) {};
		\node[fill=blue!70!white, octagon, scale=0.2] at ($(trans.east)!.9!(demod.west)+(0,.2)$) {};
		\node[fill=blue!70!white, octagon, scale=0.7] (molecule) at ($(trans.east)!.7!(demod.west)+(0,.1)$) {};
		\node[fill=blue!70!white, octagon, scale=0.5] at ($(trans.east)!.25!(demod.west)+(0,-.15)$) {};

		\draw[ultra thick, Set1-B]  (dmod.10) -- ++(.3,0) -- ++(0,-.4) -- ++(.5,0) -- ++(0,.4) -- ++(.25,0) --++(0,-.4) coordinate (c);
		\draw[ultra thick, Set1-B] (c) --       (c -| decode.west);

		\draw[ultra thick, Set1-B]  (demod.-20) .. controls ++(.3,1.8) and +(-.5,-1)  ..  (dmod.west);
		
		\node[fit=(demod) (decode), label=90:Receiver, draw, dashed, inner sep=10pt] {};
		\node[below=20pt of decode] (msg) {\verb|"message"|};
		\draw[->] (decode) -- (msg);
	\end{circuitikz}

	\caption{The three steps of the receiver: First the incoming information particles are measured (left), then the signal is demodulated (center) before being post-processed in a decoding step (right).}
	\label{fig:receiver}
\end{figure}

The first step of the receiver is to measure the amount of received information particles at a given time. In the next step, the demodulation takes place: the sensor data has to be interpreted and converted into a digital signal; a stream of (often binary) symbol values. This digital signal might then be further processed, e.g., to decode the symbol values into a message string (the authors of~\cite{Farsad2013}, for example, used their setup to send the message string \verb|"o canada"|). See Figure~\ref{fig:receiver} for an overview of the whole process.

\diff{Although simple demodulation approaches, such as threshold-based classifiers, have been used for molecular communication receivers (see \cite{Kuscu.2019} for an overview), it turns out that the demodulation actually is a very difficult task.} This is due to various sources of disturbances in the input data of the demodulation step. First of all, the physical channel (i.e., the tube) is subject to influences such as changes in flow velocity, saturation of information particles, or inaccuracies in the particle dispensing process. Also, when sending many signals in a short time, inter-symbol interference may occur, as a slow decay of information particles increases the general baseline for following measurements. Then, the measuring device can only guarantee a certain accuracy and is also influenced by the environment such as a patient wearing a medical device that uses molecular communication. The receiver from~\cite{Bartunik2020a}, for example, is susceptible to interference from magnetic fields. 

Fig.~\ref{fig:signal} shows a signal example for a transmission with amplitude values ranging from zero to five at two different symbol rates. One can see that the shapes of the peaks that encode the transmitted signal vary, especially in height; even after a normalization pre-processing step. Peaks of similar height might encode different symbols at different times within the transmission (see, for example, the peak for the first number five and the next peak for the four in Fig.~\ref{fig:signal}(a)). In Fig.~\ref{fig:signal}(b) inter-symbol interference becomes even more pronounced with the increased symbol rate.

\begin{figure}
	\centering
	\begin{tabular}[b]{c}
		\begin{tikzpicture}	
			\pgfmathsetmacro{\ypos}{1}
			\begin{axis}[
				xlabel=Time (\unit{\second}),
				ylabel=Normalised sensor signal,
				width=0.9\linewidth,
				height=0.6\linewidth,
				xmin = 0, xmax = 5,
				ymin = 0, ymax = 1
				]
				
				\addplot [Set1-B, mark=none, thick, x filter/.code={\pgfmathparse{\pgfmathresult-9.25}}, y filter/.code={\pgfmathparse{(\pgfmathresult-0)/850}}] 
				table [x=time, y=value, col sep=comma] {2Hz_rs1_6symbols_out.csv};
				
				\addplot [gray, only marks, mark=text, text mark = {0}] coordinates {(2.25, 0.1)(4.75, 0.1)};
				\addplot [gray, only marks, mark=text, text mark = {1}] coordinates {(4.25, 0.1)};
				\addplot [gray, only marks, mark=text, text mark = {2}] coordinates {(3.25, 0.1)};
				\addplot [gray, only marks, mark=text, text mark = {3}] coordinates {(1.75, 0.1)};
				\addplot [gray, only marks, mark=text, text mark = {4}] coordinates {(0.75, 0.1)(1.25, 0.1)};
				\addplot [gray, only marks, mark=text, text mark = {5}] coordinates {(0.25, 0.1)(2.75, 0.1)(3.75, 0.1)};
				
				\addplot [gray, ycomb, mark=none] coordinates {
					(0.5, 1)(1, 1)
					(1.5, 1)(2, 1)
					(2.5, 1)(3, 1)
					(3.5, 1)(4, 1)
					(4.5, 1)(5, 1)
					
				};  
			\end{axis}
		\end{tikzpicture}\\
		{{\BildSpitzmarkeFont(a):} \baselineskip=9.5pt\lineskip=0pt\BildUnterschriftFont \qty{2}{\hertz} symbol rate}
	\end{tabular}
	\begin{tabular}[b]{c}
		\begin{tikzpicture}	
			\pgfmathsetmacro{\ypos}{1}
			\begin{axis}[
				xlabel=Time (\unit{\second}),
				ylabel=Normalised sensor signal,
				width=0.9\linewidth,
				height=0.6\linewidth,
				xmin = 0, xmax = 2.5,
				ymin = 0, ymax = 1
				]
				
				\addplot [Set1-B, mark=none, thick, x filter/.code={\pgfmathparse{\pgfmathresult-1.9}}, y filter/.code={\pgfmathparse{(\pgfmathresult-0)/950}}] 
				table [x=time, y=value, col sep=comma] {4Hz_rs1_6symbols_out.csv};
				
				\addplot [gray, only marks, mark=text, text mark = {0}] coordinates {(1.125, 0.1)(2.375, 0.1)};
				\addplot [gray, only marks, mark=text, text mark = {1}] coordinates {(2.125, 0.1)};
				\addplot [gray, only marks, mark=text, text mark = {2}] coordinates {(1.625, 0.1)};
				\addplot [gray, only marks, mark=text, text mark = {3}] coordinates {(0.875, 0.1)};
				\addplot [gray, only marks, mark=text, text mark = {4}] coordinates {(0.375, 0.1)(0.625, 0.1)};
				\addplot [gray, only marks, mark=text, text mark = {5}] coordinates {(0.125, 0.1)(1.375, 0.1)(1.875, 0.1)};
				
				\addplot [gray, ycomb, mark=none] coordinates {
					(0.25, 1)(0.5, 1)(0.75, 1)(1, 1)
					(1.25, 1)(1.5, 1)(1.75, 1)(2, 1)
					(2.25, 1)(2.5, 1)(2.75, 1)(3, 1)
				};  
			\end{axis}
		\end{tikzpicture}\\
		{{\BildSpitzmarkeFont(b):} \baselineskip=9.5pt\lineskip=0pt\BildUnterschriftFont \diff{\qty{4}{\hertz} symbol rate}}
	\end{tabular}
	\caption{Normalized sensor signal for a sample transmission with different amplitude values (shown in gray) using SPIONs. The peak level indicates the transmitted value (here: numbers from zero to five). The ground truth is annotated at the peaks. As one can see, the peak heights for the numbers vary and are subject to inter-symbol interference, making demodulation difficult.}
	\label{fig:signal}
\end{figure}

This problem is well known in the MolCom literature. In~\cite{Bartunik2020}, for example, the authors try to minimize the impact of errors in the demodulation step by introducing a Gray code-based encoding. This encoding aims to minimize the distances in the binary encoding between neighbouring symbols, as error syndromes with small symbol value offsets occur more frequently than others. The demodulation step then consisted of a feature extraction step that was followed by a linear discriminant analysis (LDA). The presented method allowed for a transmission speed of \SI{4.5}{\bit\per\second}. While this approach worked well, it relied on finding Gray codes suitable for the size of the used symbol alphabet.

To address the various demodulation challenges, researchers started to investigate classification methods based on artificial intelligence as will be discussed in the next section.

\section{Artificial Intelligence for Molecular Communication}\label{sec:ai}

Many different artificial intelligence approaches have been successfully applied in the field of signal processing, e.g., in speech recognition~\cite{Deng2013} or image recognition~\cite{Krizhevsky2017}. In the domain of MolCom, the main technique used is Convolutional Neural Networks (CNNs, see, e.g.,~\cite{Sze2017} for an introductory paper).

The published papers on the demodulation process using CNNs is divided into two different approaches. One approach, the model-based approach, is to define a mathematical model of the transmission channel. Using the model, synthetic test data is generated and the demodulator/CNN is then trained on these models. The quality of such approaches highly depends on the quality of the model. An obvious shortcoming of this approach is that it is difficult or impossible to validate the results. The second approach, the data-driven approach, skips the modeling step and generates real-world data by measuring in an actual test bed. We will investigate both approaches but will focus on the data-driven approach as we consider it to produce results that are directly applicable.

\subsection{Model-based Approaches}

Early works making use of CNNs for MolCom can be found in~\cite{Yilmaz2017} and~\cite{Lee2017}. Due to the lack of an existing test bed, the authors model a diffusion-based channel. 

The authors of~\cite{Yilmaz2017} use assumptions on the physical attributes of the channel and the molecule used in the communication to derive a model of the channel. Simulation data is generated based on the model and then, in turn, used to train a convolutional neural network. The formula for the fraction of molecules arriving at the receiver at time $t$ is modeled as $$F(t) = \frac{r}{d+r}\cdot\text{erfc}\left(\frac{d}{\sqrt{4 D \cdot t}}\right),$$ where $r$ is the radius of the spherical receiver, $d$ the distance between the transmitter and receiver, $\text{erfc}$ is the complementary error function \diff{defined as $$\text{erfc}(z)=\frac{2}{\sqrt{\pi}} \int_{0}^z e^{-t^2}\text{d}t$$ (see, e.g.\ \cite[chapter 3.2]{Andrews1998} for a thorough discussion)}, and $D$ the diffusion coefficient describing the movement of the molecules.

The authors then introduce three variables to fit the model simulation data
$$
F(t, b_1, b_2, b_3) = b_i\cdot \frac{r}{d+r} \text{erfc}\left(\frac{d}{\sqrt{(4 D)^{b_2} \cdot t^{b_3}}}\right).
$$
In order to fit the parameters to simulation data, the following minimization problem is solved
$$
\argmin_{b_1, b_2, b_3} \sum_{k=1}^N \left(F(t_k, b_1, b_2, b_3) - S(t_k)\right)^2
$$
where $S(t)$ denotes simulated mean molecule arrival amounts at $N$ time instances $t$. The data set generated from this process is then used to train a CNN.

The authors of this paper only considered a single transmitter and a single receiver. This is extended in~\cite{Lee2017} to support multiple transmitters and receivers. These works are of theoretical nature as no physical experiments have been conducted to validate the approach.

A similar, model-based approach is pursued in~\cite{Qian2018}. This paper focuses even more on the mathematical model, using the more complex formula
$$F(t)=\frac{r\cdot (d - r)}{d\cdot \sqrt{4\pi D\cdot t ^3}}\cdot e^{\frac{(d-r)^2}{4D\cdot t}}.$$ Again, no physical testbed is used to validate the learned neural network. 

Unfortunately, the papers discussed in this section only present classification results visually in form of graphs, making it difficult to compare the classification rates directly.

\subsection{Data-driven Approach}
The research discussed in this section does not try to come up with a physical model of the channel but relies on measurements of MolCom in physical test beds.

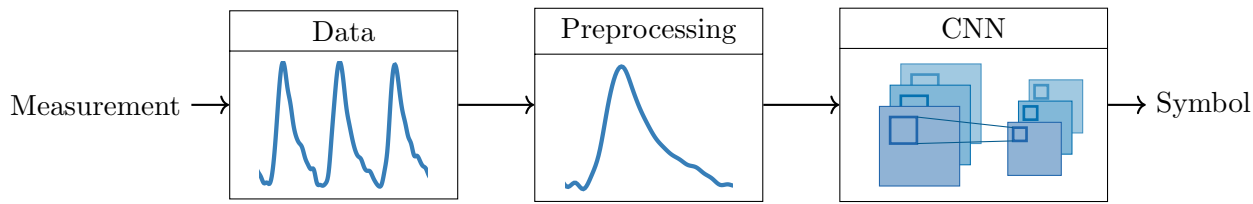
\begin{figure*}
	\centering
	\newcommand{\boxheight}{1.9}
	
	\newcommand{\mycurve}[3]{
		\begin{tikzpicture}
			\begin{axis}[
				height=\boxheight cm,
				xmin=#1, xmax=#2,
				ymin=-0.1, ymax=0.7,
				ticks=none,
				hide axis=true,
				enlargelimits=false,
				scale only axis,
				clip bounding box=upper bound,
				clip=true,
				]
				\addplot[color=Set1-B, ultra thick] table [col sep = comma, x=time, y=data]{#3};
			\end{axis}
		\end{tikzpicture}
	}
	
	\newcommand{\featurecurve}{
		\begin{tikzpicture}
			\begin{axis}[
				height=\boxheight cm,
				xmin=0.25, xmax=1.35,
				ymin=-0.1, ymax=0.7,
				ticks=none,
				hide axis=true,
				enlargelimits=false,
				scale only axis,
				clip bounding box=upper bound,
				clip=true,
				]
				\addplot[color=Set1-B, ultra thick] table [col sep = comma, x=time, y=data]{1Hz_ts1_6symbols_matlabOut_5.csv};
				
				\addplot[Set1-C, mark=*] coordinates {(0.37, 0) (0.37, 0.68)};
				\addplot[Set1-A, mark=*] coordinates {(1.28, 0.02) (1.28, 0.68)};
				\addplot[Set1-D, mark=*, only marks] coordinates {(0.62, 0.68)};
			\end{axis}
		\end{tikzpicture}
	}
	
	\newcommand{\smoothcurve}{\mycurve{0.25}{1.35}{1Hz_ts1_6symbols_matlabOut_5.csv}}
	\newcommand{\noisycurve}{\mycurve{0.4}{3.4}{1Hz_ts1_6symbols_matlabOut_DBL.csv}} 
	
	\newcommand{\cnn}{%
		\pgfdeclarelayer{bg1} 
		\pgfdeclarelayer{fg1} 
		\pgfdeclarelayer{bg2} 
		\pgfdeclarelayer{bg3} 
		\pgfdeclarelayer{fg2} 
		\pgfsetlayers{bg3,bg2,bg1,main,fg1,fg2} 
		\pgfmathsetmacro{\xshi}{2.4}
		\pgfmathsetmacro{\yshi}{0.2}
		\begin{tikzpicture}[scale=\boxheight/(2.5+0.2)]
			\clip (1,0) rectangle (4, 2.5);
			\foreach \da/\db/\offx/\offy/\cola in {fg1/main/.1/.1/RdBu-N,bg1/bg2/.3/.5/RdBu-M,bg2/bg3/.5/.9/RdBu-L}%
			{
				\begin{scope}[xshift=\offx cm,yshift=\offy cm]
					\begin{pgfonlayer}{\da}
						\path[draw=\cola, fill=\cola!50] (0,0) rectangle (1.5,1.5);
						\path[draw=\cola, line width=1] (0.2,0.8) rectangle +(0.5,0.5);
						\path[draw=\cola, line width=1] (\xshi+0.1,\yshi+0.65) rectangle +(0.25,0.25);
					\end{pgfonlayer}
					\begin{scope}[xshift=\xshi cm, yshift=\yshi cm]
						\begin{pgfonlayer}{\db}
							\path[draw=\cola, fill=\cola!50] (0,0) rectangle +(1,1);
						\end{pgfonlayer}
					\end{scope}
				\end{scope}
			}
			\begin{pgfonlayer}{fg1}
				\path[draw=PuBu-K] (0.8,0.9) -- (\xshi+0.2,\yshi+0.75);
				\path[draw=PuBu-K] (0.8,1.4) -- (\xshi+0.2,\yshi+1);
			\end{pgfonlayer}
		\end{tikzpicture}
	}
	
	\begin{tikzpicture}[struct/.style={draw, rectangle, minimum width=3cm, align=center, rectangle split, rectangle split parts=2}]
		
		\node[
		struct,
		pin={[pin edge={<-, black, thick}]left:Measurement}, 
		] (data) {Data\nodepart{second}\noisycurve};
		\node[struct, right=of data)] (preproc) {Preprocessing\nodepart{second}\smoothcurve};
		\node[
		struct, 
		anchor=east, 
		pin={[pin edge={->, black, thick}]right:Symbol}, 
		right=of preproc
		] (cnn)  {CNN\nodepart{second}\centering \cnn};
		
		\draw[->, thick] (data) --  (preproc);
		\draw[->, thick] (preproc) --  (cnn);
	\end{tikzpicture}
	\caption{The three steps of the demodulation as presented in~\cite{Bartunik2022}. Of particular interest is the last step: Application of a CNN to demodulate the signal.}
	\label{fig:cnn_overview}
\end{figure*}

One recent work using measured data is~\cite{Koo2020}. The authors use experimental data from an existing test bed that is intended for application inside a human blood vessel. They analyze a stream of single bits, with a maximal transmission rate of \qty{2}{\bit\per\second} using neural networks. Their setup is limited to the detection of binary symbols (i.\,e.\ absence/presence of information particles). The presented values for the \emph{bit error rate} of their approach using CNNs exhibit rather high error rates. Even for a transmission rate of only~\qty{0.33}{\hertz}, the error rate is still at~\qty{19}{\percent}. Due to these results, we will not further consider this publication in this section.

A more complex, i.e., larger, symbol alphabet was successfully investigated using concentration shift keying and up to 8 amplitude levels~\cite{Bartunik2022, Bartunik2022a}. These works can make use of real measurement data obtained from the concept presented in~\cite{Unterweger2018}. The physical testbed theoretically allows for a data rate of up to \SI{12}{\bit\per\second}. A high-level overview over their approach is shown in Fig.~\ref{fig:cnn_overview}. As the works~\cite{Bartunik2022a,Bartunik2022}, to the best of our knowledge, are of the few publications that consider real-world data and are not restricted to binary values, we will discuss their approach in more detail.

\subsubsection{Data Preprocessing}

For the data to be usable in a CNN, some preprocessing steps are necessary.

First of all, as the sensors used for measuring the amount of information particles have small variations in the sample rate, a linear interpolation of the data is computed. This allows to compensate for the variations.

Next, to reduce background noise, a smoothing operation using a moving average filter of width $10$ is applied.

The last preprocessing step is to split the incoming signal into sections (or intervals) that are expected to contain exactly one symbol. For this, a slope analysis is performed. As the length of a modulated symbol is approximately known due to the constant flow rate of the transport medium and the well-defined amount of particles injected for each symbol, this information is used as additional guidance in the analysis process. 

\subsubsection{CNN Topology and Training}

The authors of~\cite{Bartunik2022} make use of a one-dimensional CNN with nine layers and the PyTorch\footnote{https://pytorch.org} module for Python. 

Different weights for the neural network were trained for each of the six combinations of symbol rates and amplitude levels, i.e., symbol rates of \SI{1}{\hertz}, \SI{2}{\hertz}, and \SI{4}{\hertz} and symbol alphabets of size $6$ and $8$. Furthermore, the symbol sections of slightly varying length are interpolated to a fixed length of $128$~samples.

The neural network consists of three convolutional layers, each of which is followed by a max-pooling layer for dimensionality reduction. The last three layers are fully-connected layers and perform the final classification. The chosen number of layers is a trade-off between sufficient complexity for capturing the shapes of modulated symbols on the one hand and time necessary to train the network on the other hand. The convolutional layers use a typical kernel size, i.e. a filter size, of $7, 5$ and $3$, respectively, and stride, i.e. step size, $1$. Zero padding is used to maintain the dimensions during the convolution. For the max-pooling layers, a kernel size of $2$ is used in combination with a stride of $2$.

Dropout is a regularization method that randomly sets input values to $0$ with a probability $p$ during training. In~\cite{Bartunik2022}, dropout is used with a probability of $p=0.5$ in the first two fully-connected layers. For non-linearity, a rectified linear unit activation function after each convolutional and fully-connected layer is employed.
The details on the architecture, i.e. the individual layers, are listed in Tab.~\ref{tab:dl_architecture}.


As the loss function, \texttt{CrossEntropyLoss}, a combination of the softmax function and the cross entropy function, is used. Optimization is achieved with Adaptive Moment Estimation \cite{Kingma2017} at default initial values (initial learning rate: $10^{-3}$, $\beta_1 = 0.9$, $\beta_2 = 0.999$). 

For the training, a batch size of $64$ is used. The learning rate is decreased by a factor of $10$ when the validation set loss has not decreased for ten subsequent epochs. Once the validation loss has not decreased for $20$ subsequent epochs, training is stopped. As the used network is rather small, a training epoch only takes about one second on a Nvidia GeForce RTX 2080 graphics card. The epoch count for the training lies in the range of $40$--$100$, hence, the training is typically finished in less than two minutes.

\begin{table}
	\centering
	\caption{Architecture of CNN used in \cite{Bartunik2022}, consisting of three convolutional layers (CONV), three max-pooling layers (MAX) and three fully-connected (FC) layers. The column DO denotes whether dropout with $p=0.5$ is used. The value $C$ denotes the symbol alphabet size ($6$ or $8$).}
	\label{tab:dl_architecture}
	\setlength{\tabcolsep}{2.5pt}
	\begin{tabular}{clllccc}
		\toprule
		\textbf{Lay.} & \textbf{Type} & \textbf{Input} & \textbf{Output} & \textbf{Kernel} & \textbf{Stride} & \textbf{DO}\\
		\midrule
		1 & CONV    & $128 \times 1$    & $128 \times 64$ & 7 & 1 & \xmark\\
		2 & MAX     & $128 \times 64$   & $64 \times 64$  & 2 & 2 & \xmark\\
		3 & CONV    & $64 \times 64$    & $64 \times 128$ & 5 & 1 & \xmark\\
		4 & MAX     & $64 \times 128$   & $32 \times 128$ & 2 & 2 & \xmark\\
		5 & CONV    & $32 \times 128$   & $32 \times 256$ & 3 & 1 & \xmark\\
		6 & MAX     & $32 \times 256$   & $16 \times 256$ & 2 & 2 & \xmark\\
		7 & FC      & $16 \times 256$   & $1 \times 4096$ & - & - & \cmark\\
		8 & FC      & $1 \times 4096$   & $1 \times 4096$ & - & - & \cmark\\
		9 & FC      & $1 \times 4096$   & $1 \times C$    & - & - & \xmark\\
		\bottomrule
	\end{tabular}
\end{table}

\subsubsection{Experimental Results}
\begin{figure}[h!]
	\pgfplotstableread[header=false]{dl_4Hz_8symbols.dat}\datatable
	\pgfplotstablegetrowsof{\datatable}
	\pgfmathtruncatemacro{\numrows}{\pgfplotsretval}
	\pgfplotstablegetcolsof{\datatable}
	\pgfmathtruncatemacro{\numcols}{\pgfplotsretval}
	\xdef\LstX{}
	\xdef\LstY{}
	\xdef\LstC{}
	\foreach \Y [evaluate=\Y as \PrevY using {int(\Y-1)},count=\nY] in {1,...,\numrows}
	{\pgfmathtruncatemacro{\newY}{\numrows-\Y}
		\foreach \X  [evaluate=\X as \PrevX using {int(\X-1)},count=\nX] in {1,...,\numcols}
		{
			\ReadOutElement{\datatable}{\PrevY}{\PrevX}{\Current}
			\pgfmathtruncatemacro{\nZ}{\nX+\nY}
			\ifnum\nZ=2
			\xdef\LstX{\PrevX}
			\xdef\LstY{\PrevY}
			\xdef\LstC{\Current}
			\else
			\xdef\LstX{\LstX,\PrevX}
			\xdef\LstY{\LstY,\PrevY}
			\xdef\LstC{\LstC,\Current}
			\fi
		}
	}
	\edef\temp{\noexpand\pgfplotstableset{
			create on use/x/.style={create col/set list={\LstX}},
			create on use/y/.style={create col/set list={\LstY}},
			create on use/color/.style={create col/set list={\LstC}},}}
	\temp
	\pgfmathtruncatemacro{\strangenum}{\numrows*\numcols}
	\pgfplotstablenew[columns={x,y,color}]{\strangenum}\strangetable
	
	\centering
	\begin{tikzpicture}
		\begin{axis}[%
			tick align=outside,
			minor tick num=0,
			xlabel=Demodulated symbol,
			xlabel near ticks,
			xmin=-0.5, xmax=7.5,
			xtick={0, 1, ..., 7},
			ylabel=Transmitted symbol,
			ymin=-0.5, ymax=7.5,
			ytick={0, 1, ..., 7},
			height=\linewidth,
			width=\linewidth,
			point meta min=0,
			point meta max=1,
			point meta=explicit,
			scale mode=scale uniformly,
			mesh/cols=8,
			font=\footnotesize,
			nodes near coords black white/.style={
				small value/.style={
					text=black,
				},
				large value/.style={
					text=white,
				},
				every node near coord/.style={
					check for zero/.code={
						\pgfmathfloatifflags{\pgfplotspointmeta}{0}{
						}{
							\begingroup
							\pgfkeys{/pgf/fpu}
							\pgfmathparse{\pgfplotspointmeta<#1}
							\global\let\result=\pgfmathresult
							\endgroup
							%
							%
							\pgfmathfloatcreate{1}{1.0}{0}
							\let\ONE=\pgfmathresult
							\ifx\result\ONE
							\pgfkeysalso{/pgfplots/small value}
							\else
							\pgfkeysalso{/pgfplots/large value}
							\fi
						}
					},
					check for zero,
				},
			},
			nodes near coords black white=\whitethreshold,
			]
			\addplot [
			matrix plot,
			nodes near coords,
			nodes near coords style={anchor=center},
			nodes near coords style={/pgf/number format/.cd, fixed, precision=2, zerofill, skip 0.},
			point meta=explicit,
			] table [meta=color,col sep=comma] \strangetable;
		\end{axis}
	\end{tikzpicture}
	\caption{Classification performance of the CNN using an alphabet of 8 symbol values and a symbol rate ot \qty{4}{\hertz}. The classification rate is at \qty{49}{\percent}. One can see that most misclassifications do not introduce large changes in the symbol, e.g., there are very few cases of a transmitted symbol~$6$ being classified as~$4$.}
	\label{fig:classification_performance}
\end{figure}
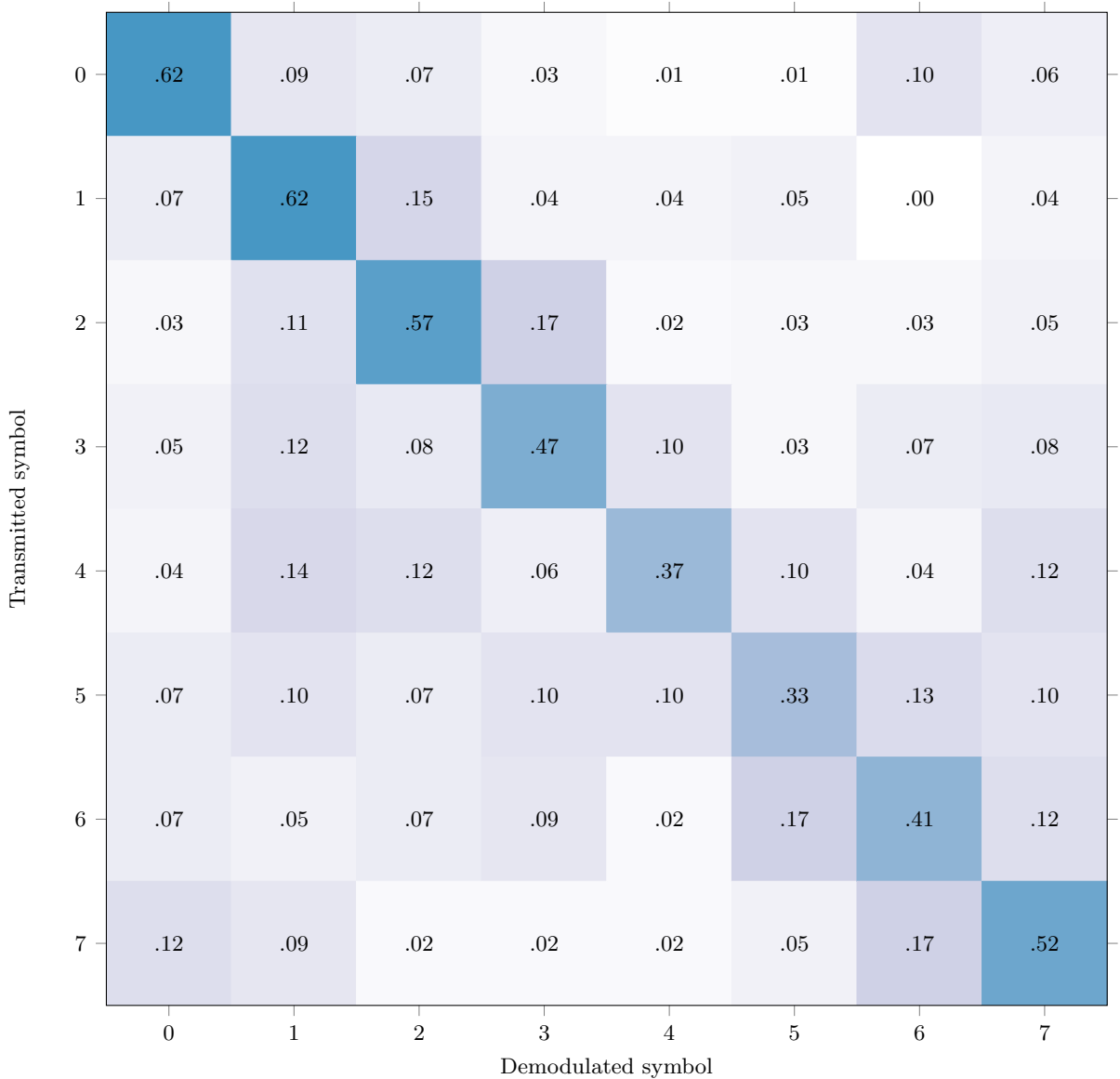

The classification performance of the CNN can be visualized using a confusion matrix, comparing the demodulated symbol value to the actually transmitted symbol value. Figure~\ref{fig:classification_performance} shows the confusion matrix for the most complex scenario of sending $8$ different symbols with a symbol rate of~\qty{4}{\hertz}. The shown values are probabilities for the occurrence of a specific classification outcome, dependant on the actually transmitted symbol.

While the classification results for this difficult case still are in a reasonable range, it is worth discussing the observed problems in this scenario.
At a symbol rate of \qty{4}{\hertz} the classification accuracy deteriorates due to strongly increased inter-symbol interference and an increased significance of timing errors. 
However, for the scenario with a symbol alphabet size of $6$ an average of \qty{65}{\percent} correctly classified symbols is still achieved at a symbol rate of~\qty{4}{\hertz}. Tab.~\ref{tab:comp} lists classification offset for each combination of symbol alphabet size and symbol rate.  As one can see, using a symbol rate of~\qty{2}{\hertz} is already sufficient for a very reliable demodulation for both, alphabets with $6$ and $8$ symbols. 
	The results for a symbol rate of~\qty{1}{\hertz} are, counter-intuitively, slightly worse than for \qty{2}{\hertz}. This is due to time-dependant variations of the injection behaviour at the transmitter causing a lower accuracy for longer delays between symbols (a detailled discussion of these effects can be found in \cite{Bartunik2023}).

The noisy-channel coding theorem~\cite{Shannon1948} provides an upper boundary for the information rate of a transmission channel, given a tolerable remaining bit error rate $p_\text{b}$. The maximally achievable net data rate, given a gross data rate $R_\text{g}$ and a binary channel with noise level (i.\,e., bit error rate) $f$, is
$$
R = R_\text{g} \frac{1-H_2\left(f\right)}{1-H_2\left(p_\text{b}\right)}
$$
with the binary entropy function
$$
H_2\left(x\right) = x \log_2 \left( \frac{1}{x}\right) + \left(1-x\right) \log_2\left(\frac{1}{1-x}\right).
$$

With a bit error rate of $p_\text{b} = \qty{1}{\percent}$ the achievable effective data rate is more than \qty{5.5}{\bit\per\second} for the combination of $8$~symbol values and a symbol rate of \qty{2}{\hertz}.

\begin{table}[thb]
	\centering
	\caption{Comparison between the demodulation for different symbol alphabets and at different symbol frequencies. The shown values are probabilities for the occurrence of a specific demodulation offset (i.\,e., mis-classification) from the transmitted value.}
	\label{tab:comp}
	\begin{tabular}{cllllll}
		\toprule
		& \multicolumn{3}{c}{\textbf{6 Symbols}} & \multicolumn{3}{c}{\textbf{8 Symbols}} \\
		\cmidrule(lr){2-4}
		\cmidrule(lr){5-7}
		\textbf{Offset} & \textbf{\SI{1}{\hertz}} & \textbf{\SI{2}{\hertz}} & \textbf{\SI{4}{\hertz}} & \textbf{\SI{1}{\hertz}} & \textbf{\SI{2}{\hertz}} & \textbf{\SI{4}{\hertz}} \\
		\midrule
		0 & 0.93 & 0.99 & 0.65 & 0.95 & 0.98 & 0.49\\
		1 & 0.06 & 0.01 & 0.15 & 0.04 & 0.02 & 0.20\\
		2 & 0.00 & 0.00 & 0.09 & 0.00 & 0.00 & 0.09\\
		3 & 0.00 & 0.00 & 0.05 & 0.00 & 0.00 & 0.08\\
		4 & 0.00 & 0.00 & 0.03 & 0.00 & 0.00 & 0.05\\
		5 & 0.00 & 0.00 & 0.04 & 0.00 & 0.00  & 0.03\\
		6 & n/a & n/a & n/a & 0.00 & 0.00 & 0.04\\
		7 & n/a & n/a & n/a & 0.00 & 0.00 & 0.02\\
		\bottomrule
	\end{tabular}
\end{table}

So far, none of the published research papers adequately deal with the dynamic time-varying nature (e.g., memory in the transmission channel). A potential solution for this issue might be to use neural network implementations that are capable of keeping track of previous states. For this, Recurrent Neural Networks (RNNs,~\cite{Rumelhart1986}), more precisely Long Short-Term Memory (LSTM) networks~\cite{Hochreiter1997} are promising candidates to address the issue. To the best of our knowledge, no research into this direction using experimental data has been conducted so far.

\section{Conclusion}
The paper gave a brief introduction into the field of molecular communication. Setups using either air or fluid as a transmission medium have been presented and their advantages and drawbacks have been discussed, e.g., with respect to the possible field of application.

Afterwards, the most difficult task in MolCom, at least from a computer science perspective, has been identified: the demodulation of the measured molecules/signal. We showed that artificial intelligence, more precisely, convolutional neural networks, are an efficient technique for demodulation that is already successfully employed. However, it should be noted that many of the publications do not work on measured data but on theoretical models. The results presented in this paper have been obtained using actual measurements on a self-developed and self-built testbed.

Experimental research for time-varying signals in MolCom is scarce, especially regarding the use of artificial intelligence. 
An approach to tackle this issue could be to perform the training and inference steps concurrently in an online fashion. Such a system could then accommodate for fluctuations in the channel properties. Depending on the application scenario, however, such an online approach might have power requirements that are prohibitively high. Demodulation of time-varying MolCom signals is still an open research question and is left for future work.

\section*{Acknowledgments}
This work was supported in part by the German Federal Ministry of Education and Research (BMBF), project 
MAMOKO (16KIS0913K)

\bibliographystyle{plain}
\bibliography{generated,header,vonJens,vonMax}


\begin{minipage}{\linewidth}
	

	\vrule height.5pt depth 0pt width\linewidth
	\newline
	\begin{wrapfigure}[6]{l}[0pt]{25mm}
		\vtop{%
			\vskip-6ex
			\hbox{%
				\includegraphics[width=1in]{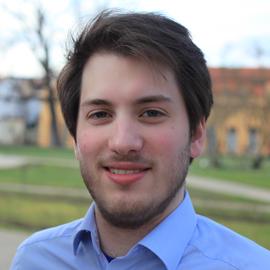}%
			}%
		}
	\end{wrapfigure}

	\vspace{-2.5em}
	\authorone
	
\end{minipage}


\bigskip
\begin{minipage}{\linewidth}
	
	\vrule height.5pt depth 0pt width\linewidth
	\newline
	\begin{wrapfigure}[9]{l}[0pt]{25mm}
		\vtop{%
			\vskip-6ex
			\hbox{%
				\includegraphics[width=1in]{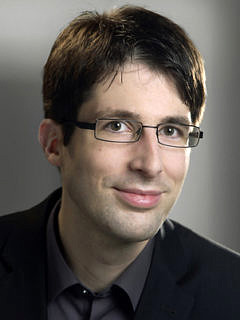}%
			}%
		}
	\end{wrapfigure}

	\vspace{-2.5em}
	\authortwo
	
\end{minipage}



\bigskip
\begin{minipage}{\linewidth}
	
	\vrule height.5pt depth 0pt width\linewidth
	\newline
	\begin{wrapfigure}[9]{l}[0pt]{25mm}
		\vtop{%
			\vskip-6ex
			\hbox{%
				\includegraphics[width=1in]{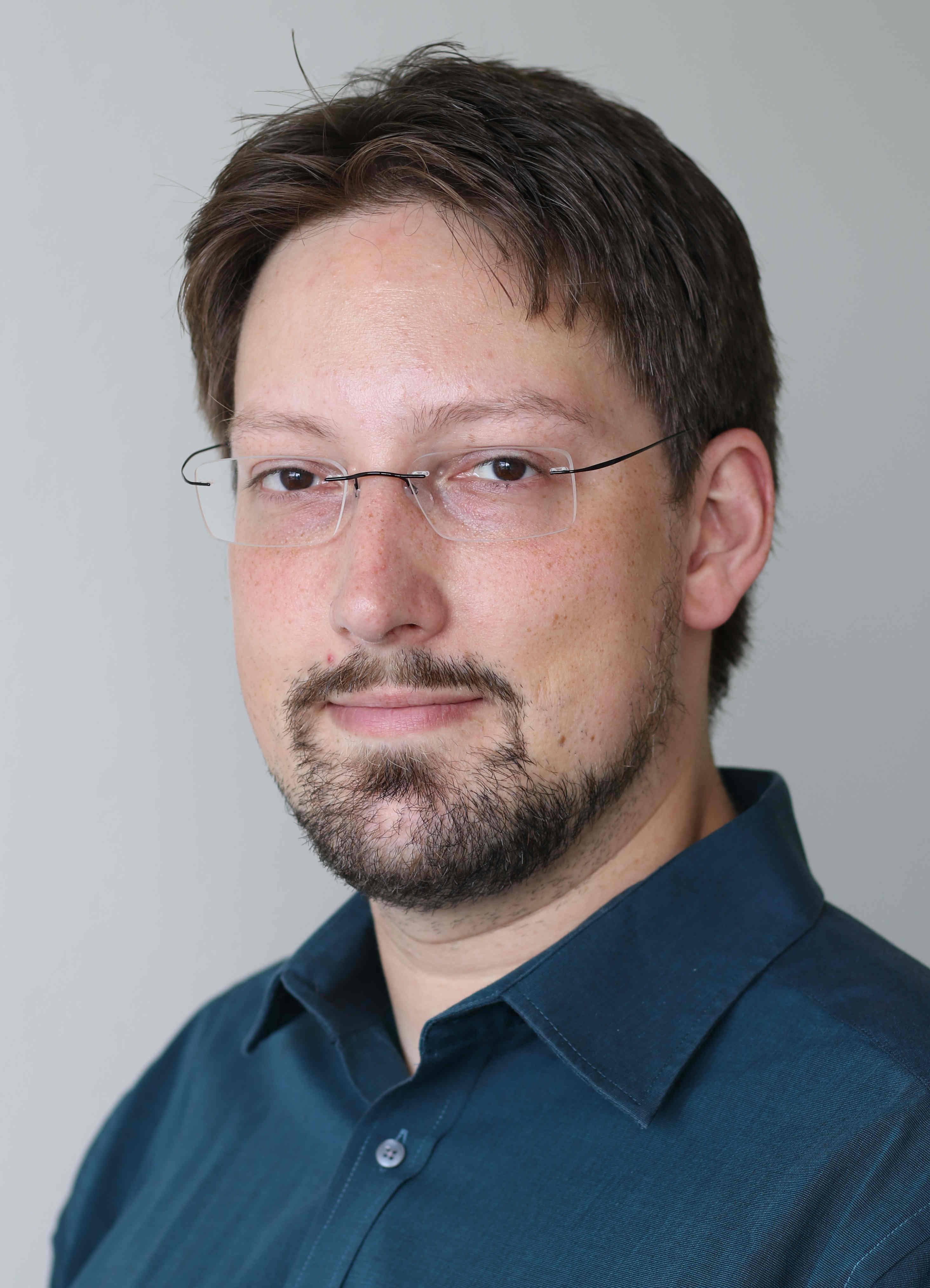}%
			}%
		}
	\end{wrapfigure}

	\vspace{-2.5em}
	\authorthree
	
\end{minipage}

\end{document}